\documentclass{article}

\usepackage{arxiv}

\usepackage[utf8]{inputenc} 
\usepackage[T1]{fontenc}    
\usepackage{hyperref}       
\usepackage{url}            
\usepackage{booktabs}       
\usepackage{amsmath, amssymb, amsfonts}       
\usepackage{nicefrac}       
\usepackage{microtype}      
\usepackage{lipsum}		
\usepackage{graphicx}
\usepackage{natbib}
\usepackage{doi}

\title{Aerodynamic characterization of two tandem wind turbines under yaw misalignment control using actuator line model}

\author{ Yu Tu \\
	School of Naval Architecture, Ocean and Civil Engineering\\
	Shanghai Jiao Tong University\\
	Shanghai 200240, China \\
	\And
        \href{https://orcid.org/0000-0001-6097-7217}
	{\includegraphics[scale=0.06]{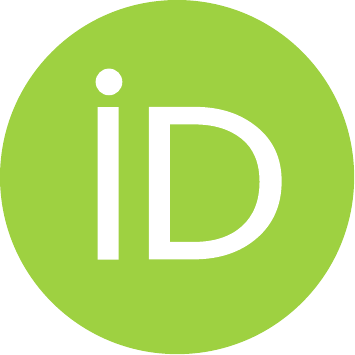}\hspace{1mm}Kai Zhang} \\
	School of Naval Architecture, Ocean and Civil Engineering\\
	Shanghai Jiao Tong University\\
	Shanghai 200240, China \\
	\texttt{kai.zhang@sjtu.edu.cn} \\
	\AND
	Zhaolong Han \\
	School of Naval Architecture, Ocean and Civil Engineering, Shanghai Jiao Tong University \\
	Shanghai 200240, China \\
	\And
	Dai Zhou \\
	School of Naval Architecture, Ocean and Civil Engineering, Shanghai Jiao Tong University \\
	Shanghai 200240, China \\
	\And
	Onur Bilgen \\
	Department of Mechanical and Aerospace Engineering, Rutgers University \\
	Piscataway, 08854, New Jersey, USA\\
}

\date{February 3,2023}

\renewcommand{\shorttitle}{Effects of yaw misalignment control on the aerodynamics of two tandem turbines}

\begin{document}
\maketitle

\begin{abstract}
Yaw control has proven to be promising in alleviating the wake effects that plague the efficiency of wind farms.
In this work, the actuator line modeling (ALM) method is adopted to simulate the flows over two tandem turbines distanced by $3 - 7$ rotor diameters, with the yaw angle of the upstream rotor varying from $\gamma_1=0^{\circ}$ to $50^{\circ}$.
The aim is to provide a comprehensive aerodynamic characterization of this simple wind farm under yaw misalignment control.
With increasing yaw angle, the power generated by the downstream rotor increases, compensating the power loss in the upstream rotor, and resulting in significantly higher total power of the two turbines than that without yaw control.
The maximum power output is achieved as the upstream wake of the yawed rotor is redirected away from the downstream rotor plane.
Behind the downstream rotor, the secondary steering phenomenon is observed, where the wake is also redirected from the centerline.
The use of the actuator line model also reveal unsteady aerodynamic characteristics that can not be captured by lower-fidelity models.
For the upstream rotor, the yaw misalignment results in time-varying change in the local angle of attack on the blade, giving rise to unsteady loading.
The downstream rotor is partially submerged in the deflected wake incurred by the yawed upstream rotor.
As the blade revolves into and out of the wake deficit, the blade experiences cyclic loading, leading to even stronger fluctuations in the aerodynamic loads than the upstream rotor.
These analysis provides a comprehensive understanding of the yaw control effects on the two tandem rotors from the perspectives of aerodynamic performance, wake profiles, and unsteady characteristics.
The insights gained from the present study can aid the design of collective yaw control strategies of wind farms, and lay the foundation for assessing the fatigue damage associated with yaw misalignment.
\end{abstract}


\section{Introduction}
\label{sec:intro}

Over the years, wind power has proven to be a viable alternative to the conventional fossil-based energy sources \citep{veers2019grand}. 
The offshore environment is more attractive for wind energy development because the air flow is typically stronger and more consistent compared to onshore flow.
However, the high cost of energy still presents an urgent problem for the deployment of large-scale offshore wind farms.
Specifically, the efficiency and service life of offshore wind farm are plagued by the wake interactions between turbines \citep{VERMEER2003467,troldborg2011numerical,sun2020review}. 
As the incoming wind passes through the upstream turbine, the wake forms with reduced wind speed and increased turbulence intensity. 
Submerged in the wakes, the downstream turbines not only generate less power but also suffer from higher fatigue loading.
According to \citet{barthelmie2009modelling}, the annual average loss caused by wake effects in large-scale wind farms accounts for about $10\% - 20\%$ of the total power generation. 

Wind farm control is a new area of research that has rapidly become a key enabler for the development of large wind farm projects \citep{andersson2021wind}.
Conventional wind farm control strategy seeks maximizing the energy output of each individual turbines.
However, such strategy does not take the entirety of wind farm as well as the spatial correlations into account. 
On the other hand, the holistic wake control strategy aims to improve the overall performance of wind farm by operating some wind turbines at sub-optimum conditions \citep{wagenaar2012controlling,andersson2021wind,houck2022review,meyers2022wind}.
Among various wind farm control strategies, wake steering via yaw control emerges as particularly effective.
In traditional way, the yaw system of wind turbine needs to track the variable ambient wind direction \citep{chen20212} and turn the rotor to align with it.
Unlike that, the yaw control strategy is realized by applying intentional yaw misalignment between the incoming flow and upstream turbines, which induces reduced power output on these units but direct the velocity deficit away from the downstream turbines.

The implementation of wake steering for wind farm requires thorough understanding of the wake dynamics associated with yawed turbines.
The most notable feature of a yawed turbine wake is arguably the formation of a counter-rotating vortex pair \citep{howland2016wake,bastankhah2016experimental,fleming2018}, which resembles the wingtip vortices in the wake of finite-aspect-ratio wings \citep{anderson2011fundamental,zhang2020formation}.
The resulting velocity deficit in the wake of the yawed turbine takes the form of a curled kidney-like shape.
In recent years, several advanced wake models have been proposed to depict the highly three-dimensional flow features behind the yawed turbines.
Recognizing the similarity of the wakes behind the yawed turbine and finite-span wings, \citet{shapiro2018modelling} treated the yawed rotor disk as a lifting body, and the Prandtl lifting line approach is used to correlate the transverse velocity induced by the vortex system with the lateral thrust force.
\citet{zong2020point} expressed the rates of vorticity shedding at rotor blade tips using vortex cylinder theory to determine the trailing vorticity distribution behind a yawed rotor.
\citet{bastankhah2022vortex} considered the wake edge of the yawed turbine as an ideally thin vortex sheet, and solved the vortex sheet equation for evolution in the wake.
Based on these calculations, they modified the Gaussian wake model by incorporating the predicted shape and deflection of the curled wake to predict the wake profiles behind yawed turbines.
\citet{king2021control} developed the Gaussian curl hybrid (GCH) wake model, which is able to reproduce the secondary effects of wake steering in large arrays of turbines.
This model is implanted into the open-source wind farm optimization toolbox FLORIS \citep{FLORIS_2021}.
These analytical models reveal useful insights into the flow physics of yawed turbines, and are essential in designing of wake mitigation strategies for wind farms.

The efficacy of yaw control in improving the wind farm efficiency has been demonstrated in a number of experimental studies.
\citet{adaramola2011experimental} studied yaw control of two aligned wind turbines with a streamwise spacing of four rotor diameters through wind tunnel measurements, and found that yaw control can enhance the total power generation by 12\%.
\citet{Campagnolo_2016} experimentally studied the effects of yaw misalignment on three turbines. 
It was revealed that yawing the front row by $20^{\circ}$ and second row by $16^{\circ}$ improved the total farm power production by 15\%.
\citet{bastankhah2019wind} performed wind tunnel experiments to study the performance of a model wind farm with five turbine rows under a wide variety of yaw angle distributions.
Their results showed that yaw angle control can increase the overall wind farm efficiency as much as 17\% with respect to fully non-yawed conditions. The most successful yaw angle distributions were found to be those with a relatively large yaw angle value for the first turbine row, and then, the yaw angle decreases progressively for downwind rows until it eventually becomes zero for the last one.
\citet{bartl2018wind_b} studied the effects of intentional yaw misalignment on the power production and loads of a downstream turbine are investigated for full and partial wake overlap. It shows that the increase in combined power is at the expense of increased yaw moments on both the upstream and downstream turbines.
\citet{aju4194363influence} performed systematic wind tunnel experiments to quantify the power output fluctuations and unsteady aerodynamic loads of model wind farms with three rows and three columns across various yaw angles.
Apart from these wind tunnel experiments, wake steering via yaw control has also been performed in a number of field tests \citep{fleming2017full,howland2019wind,simley2021results,howland2022collective}.
These studies have shown the great potential of wake steering in enhancing the power generation efficiency of commercial wind farms in realistic operating conditions.

Computational fluid dynamics (CFD) methods have also been instrumental in bridging the gap between analytical wake models and wind tunnel and field tests.
Due to the high computational cost in resolving the boundary layers of the rotating blades at high Reynolds numbers \citep{mittal2016blade,lawson2019blade,de2022blade,miao2017investigation}, most of the studies have used simplified turbine models.
The actuator disk model (ADM) is the simplest rotor modeling technique in CFD applications. 
The blade swept area forms a full disc, which represents the rotor.
The most basic form of ADM assumes uniform load distribution on the disc.
Thus, it is necessary to provide the aerodynamic coefficients as input to the model, which can be nontrivial in yawed conditions \citep{hur2019review,howland2020optimal,howland2022optimal,heck2022modeling}.
In a more advanced version of ADM, the blade element theory is incorporated into the model, allowing it to account for the radial distribution of the loads as well as the tangential forces.
\citet{lin2022large} showed that this version of ADM yields flow statistics that are in better agreement with the wind-tunnel measurements for both nonyawed and yawed conditions compared to the traditional ADM.
The actuator line model (ALM) developed by \citet{shen2005tip} computes the turbine-induced forces on line elements distributed on the moving turbine blades, which introduced temporal dependency to the turbine model.
Thus, ALM is well-suited for studying the unsteady aerodynamics of wind turbines, and has been the \emph{de facto} tool for state-of-the-art wind farm simulation \citep{stevens2017flow,stevens2018comparison,shapiro2022turbulence}. 

The objective of this paper is to understand the effects of yaw misalignment on the aerodynamics of two tandem turbines.
Although this simple wind farm layout has been investigated extensively in the literature as reviewed above, the wake profiles and the unsteady flow physics associated with the yaw control has not been elucidated in detail.
In addition, previous studies have mostly employed relatively complex setups involving atmosphere boundary layer, wind shear, inlet turbulence, etc., which could hinder a clear comprehension of the yaw misalignment effects.
To address the above questions, this paper presents large-eddy simulations with actuator line model to simulate the flow over two tandem wind turbines, and characterize the aerodynamic performance, wake profiles, and unsteady flow physics over a wide range of yaw angles.
We assume the flow to be uniform and void of ground effects to isolate the effects of yaw misalignment among other factors.
In what follows, we introduce the numerical methods, computational setup and validation in section \ref{sec:method}.
The results are presented in section \ref{sec:results}, in which we discuss the aerodynamic performance, wake profiles, and unsteady characteristics of the two tandem rotors.
We conclude this paper by summarizing our findings in section \ref{sec:conclusions}.

\section{Computational setup}
\label{sec:method}

\subsection{Problem description}
\label{sec:description}
In this study, a 5 MW reference turbine designed by \citet{jonkman2009definition} at the National Renewable Energy Laboratory (referred to as NREL 5 MW turbine hereafter) is used as the model rotor. 
This horizontal-axis, upwind turbine has three blades with a rotor diameter of $D=126$ m.
The rated wind speed is 11.4 m/s.
The turbine hub, tower and ground effects are not considered in this study.

\begin{figure}
    \centering
    \includegraphics[width=0.6\textwidth]{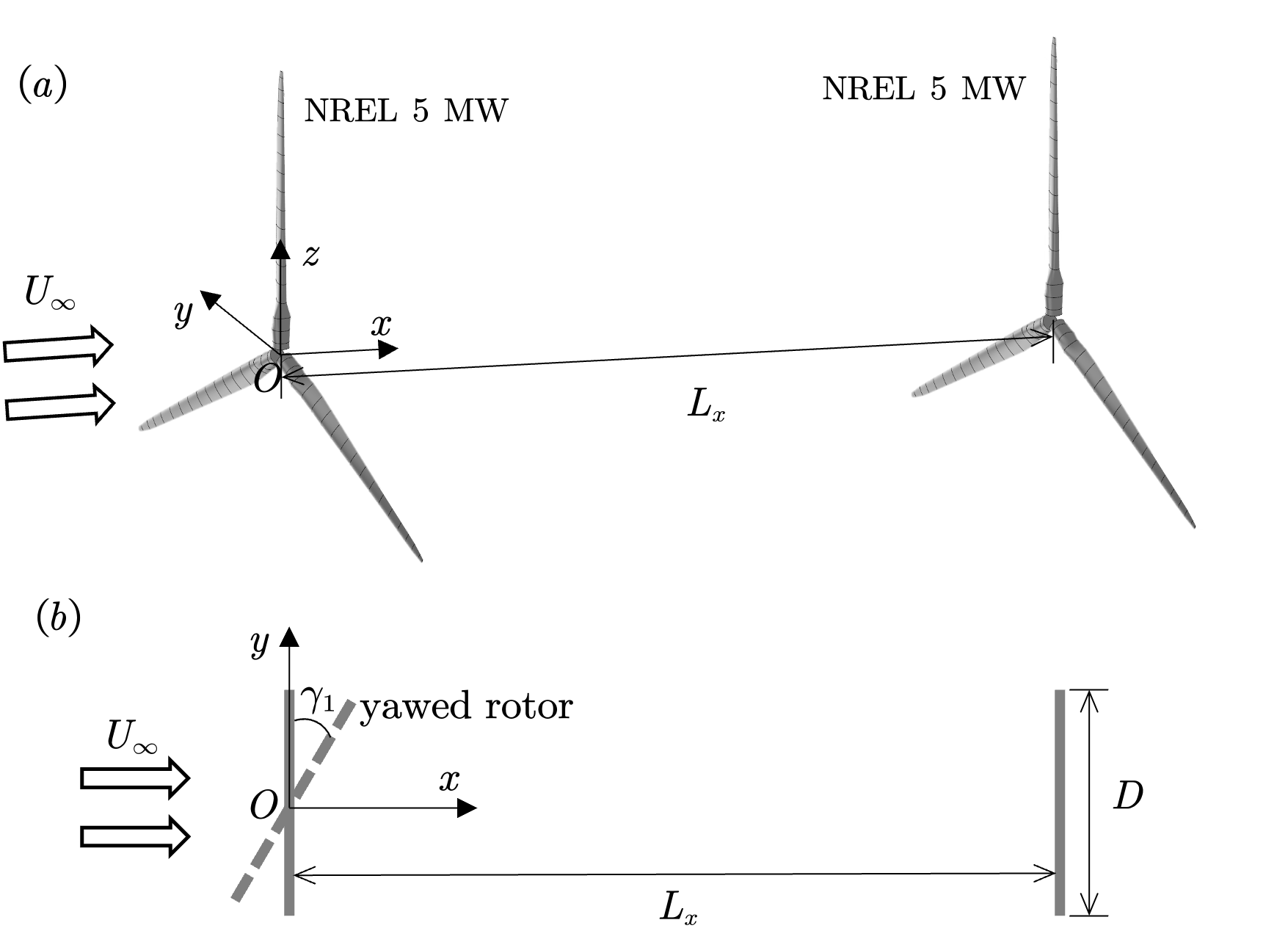}
    \caption{Description of the problem. ($a$) Perspective view of the two turbines and ($b$) $x$-$y$ plane view. The thick gray lines in $(b)$ represent the rotor planes.}
    \label{fig:description}
\end{figure}

Two NREL 5 MW rotors distanced by $L_x$ are placed in tandem along the streamwise direction, as shown in figure \ref{fig:description}.
These two rotors are subjected to uniform inflow velocity, which is set to the rated wind speed $U_{\infty}=11.4$ m/s.
Since the current study focuses on the aerodynamic characterization of the two rotors, we ignore the rotation speed control and set the tip speed ratios (defined as $\lambda=\Omega D/(2U_{\infty})$, where $\Omega$ is the angular velocity) of both turbines to be 8, unless otherwise stated. 
The yaw control is actuated around the $z$ axis of the upstream rotor, while the downstream rotor remains perpendicular to the freestream.
We characterize the aerodynamics of this simple wind farm configuration with the upstream yaw angle varying from $\gamma_1=0^{\circ} - 50^{\circ}$.

\subsection{Numerical methods}
\label{sec:numerical}
We use large eddy simulation (LES) to solve for the flows over two tandem turbines.
The filtered incompressible Navier-Stokes equation reads as 
\begin{equation}
\frac{\partial \tilde{u}_i}{\partial x_i} =0, \\
\frac{\partial \tilde{u}_i}{\partial t} + \tilde{u}_j\frac{\partial \tilde{u}_i}{\partial x_j} = -\frac{1}{\rho}\frac{\partial \tilde{p}}{\partial x_i} + \nu\frac{\partial}{\partial x_j}\frac{\partial \tilde{u}_i}{\partial x_j} + \frac{\partial \tau_{ij}}{\partial x_j} + f_i,
\label{equ:NS}
\end{equation}

where $u_i$ and $p$ are the velocity and pressure, $\tilde{\cdot}$ represents the resolved flow quantities, and $\rho$ and $\nu$ are the air density and kinematic viscosity, respectively.
$\tau_{ij}$ is the subgrid-scale stress (SGS) tensor, which is expressed according to the Boussinesq approximation with the introduction of a turbulent eddy viscosity $\nu_t$
\begin{equation}
\tau_{ij}=\frac{2}{3}k_t\delta_{ij}-2\nu_t\tilde{S}_{ij}.
\end{equation}
here, $k_t=\tau_{kk}/2$ is the SGS turbulent kinetic energy and $\tilde{S}_{ij}=\left({\partial \tilde{u}_i}/{\partial x_j}+{\partial \tilde{u}_j}/{\partial x_i} \right)/2$ is the rate of strain tensor computed from the resolved scales. The $k$-equation model is selected to calculate the kinematic energy $k_t$: 
\begin{equation}
\frac{\partial k_t}{\partial t}+\frac{\partial}{\partial x_j}\left(\tilde{u}_jk_t\right)=2\nu_t\tilde{S}_{ij}\tilde{S}_{ij}+\frac{\partial}{\partial x_j} \left[\left(\nu+\nu_t \right)\frac{\partial k_t}{\partial x_j} \right] -C_{\epsilon}k_t^{1.5}\Delta^{-1},
\end{equation}
where the SGS viscosity is given by $\nu_t=C_k\Delta k_t^{0.5}$. The model coefficients $C_{\epsilon}$ and $C_k$ are dynamically computed as part of the solution based on the Germano identity \citep{germano1991dynamic} with test filter $\widehat{\Delta}=2\Delta$ ($\Delta=\sqrt{\Delta_x\Delta_y\Delta_z}$ is the nominal grid size) by the least square minimization procedure proposed by \citet{lilly1992proposed}.

\begin{figure}
\centering
\includegraphics[width=1.0\textwidth]{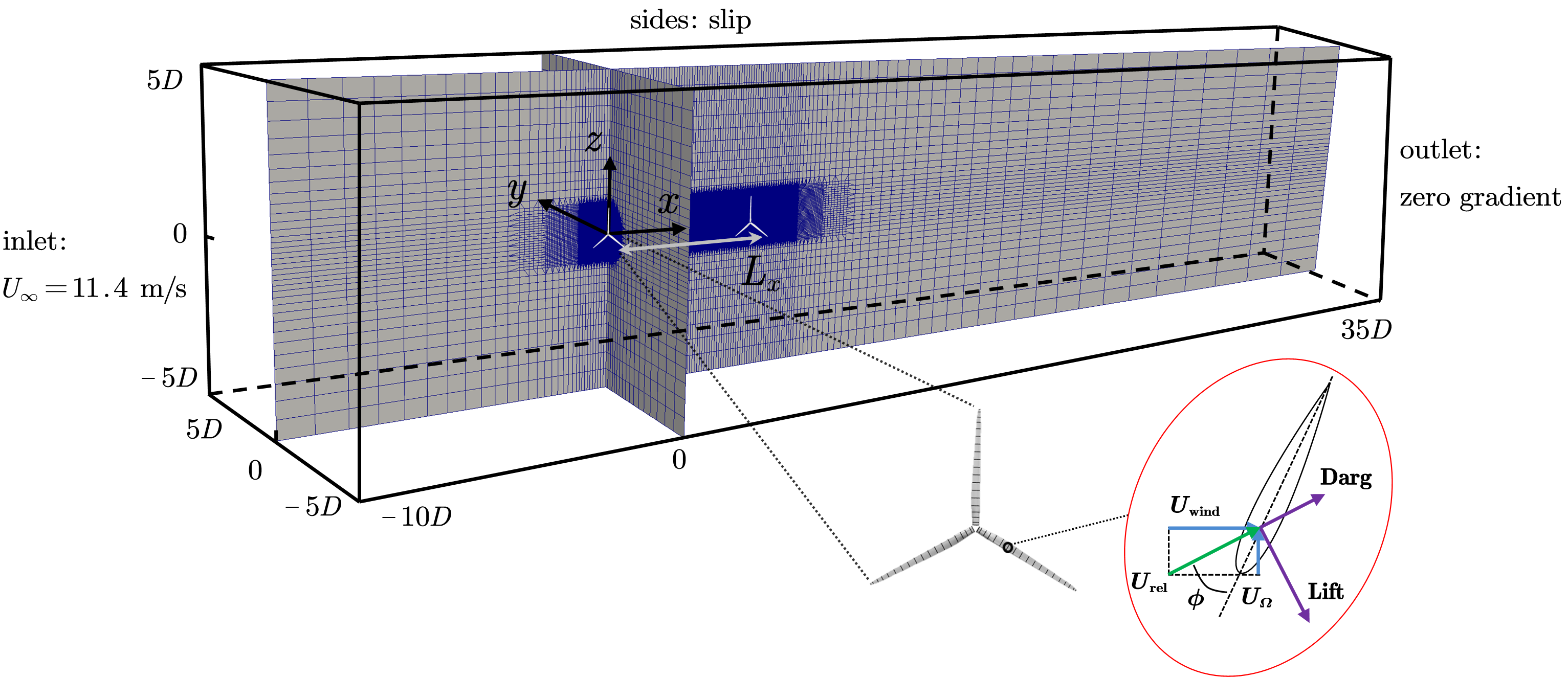}
\caption{\label{fig:grid}Computational setup and illustration of ALM.}
\end{figure}

The $f_i$ term in equation (\ref{equ:NS}) represents the body force imposed by the wind turbine. 
The actuator line model (ALM) is used to calculate these body forces.
As shown in figure \ref{fig:grid}, the ALM discretizes the blade into a series of 2D airfoil sections along the radial direction, and the point at the quarter chord point of each section is called the actuation point.
The two-dimensional sectional lift and drag forces are calculated as
\begin{equation}
f_l=\frac{1}{2}C_l \rho c U_{rel}^2, \quad f_d = \frac{1}{2}C_d\rho c U_{rel}^2,
\end{equation}
where $c$ is the chord length of the local airfoil, and $U_{rel}=\sqrt{U_{\Omega}^2+U_{\textrm{wind}}^2}$ is the local wind velocity relative to the blade ($U_{\Omega}$ and $U_{\textrm{wind}}$ are the velocity of blade rotation and wind velocity at the actuator point, respectively).
$C_l$ and $C_d$ are the lift and drag coefficient of local 2D airfoil profile, which is precalculated and tabulated with respect to the angle of attack. 
The local angle of attack $\phi$ of the 2D airfoil section is taken to be the angle between the chord and the velocity at the actuator segment. 
To account for the tip effects, the tip correction factor proposed in \citet{shen2005tip} is implemented in the ALM.
The lift and drag forces are projected into the flow field by taking the convolution with a 3D Gaussian kernel $\eta$ for each blade element
\begin{equation}
\eta = \frac{1}{\epsilon^3\pi^{3/2}}\exp(-(d/\epsilon)^2),
\end{equation}
where $d$ is the distance between the measured point and the actuator point on the blade. $\epsilon$ is the Gaussian width that determines the concentration of the distributed load, and is set as twice the local grid size, as suggested by \citet{troldborg2009actuator}.

The two tandem NREL 5 MW rotors are placed in a rectangular computational domain, which covers $(x,y,z)\in [-10D, 35D]\times[-5D, 5D]\times[-5D, 5D]$, as shown in figure \ref{fig:grid}.
The resulting blockage ratio is $0.78\%$.
The center of the upstream rotor is placed at $(x,y,z)=(0,0,0)$, and the downstream one at $(x,y,z)=(L_x,0,0)$, where $L_x$ is the spacing between the two rotors.
The actuator line representing the turbine blade is discretized into 19 segments along the span.

The flows over the turbines are simulated using the solver pimpleFoam from the open-source CFD toolbox OpenFOAM \citep{weller1998tensorial}.
We use the actuator line model implemented in the turbinesFoam library \citep{bachant2016actuator}, which has seen application in a number of studies \citep{zhang2020multi,onel2021investigation,liu2022evaluating}.
Uniform mesh is used to resolve the flows near and in the wake of the rotors.
A detailed mesh dependency test is presented in section \ref{sec:validation}.
The simulations employ a fixed Courant number of $CFL_{\max}=0.1$, as suggested by \citet{troldborg2009actuator}.
At the inlet boundary, a uniform velocity at the rated wind speed $U_{\infty}=11.4$ m/s is prescribed.
A zero-gradient condition is applied to the velocity at outlet, where a reference pressure $p_{\infty}=0$ is specified.
The rest of the boundaries are set as slip.

The FLORIS v3.0 code (FLOw Redirection and Induction in Steady State, \citet{FLORIS_2021}), which is a control-oriented wind farm simulation software, is used as a low-fidelity reference to the ALM results in this study.
FLORIS incorporates several steady-state engineering wake models to account for the wake interaction effects between turbines, and is widely accepted in wind farm control and layout studies \citep{doekemeijer2019tutorial,gebraad2017maximization}.
We use the the Gaussian curl hybrid (GCH) wake model \citep{king2021control}, which better reproduces the secondary effects of yawed wake, including the yaw-added wake recovery as well as the secondary wake steering.

By default, FLORIS calculates the aerodynamic performance of the turbine using the input table which contains the steady-state responses as a function of inflow velocity defined in \citet{jonkman2009definition}.
This approach implicitly dictates the tip speed ratio based on the upstream velocity, suggesting that the downstream rotor operates at lower rotational speed to maintain optimum tip speed ratio (based on the wake velocity of the upstream rotor).
This is different from the settings for ALM, in which both rotors operates at the same rotational speed as mentioned in section \ref{sec:description}.
To ensure a fair comparison of the results obtained from the two models, the input table of FLORIS is modified to tabulate power and thrust coefficients (precalculated by ALM) under the same rotational speed regardless of the inflow velocity.
In addition, both the wind shear and the turbulence intensity are set as zero in FLORIS.
The hub height is modified to be high enough to avoid the ground effect. 
The other parameters are set as default.

\subsection{Validation}
\label{sec:validation}

\begin{table}
\centering
\caption{\label{tab:turbine}Aerodynamic parameters of a single turbine at three grid sizes. The tested tip speed ratio is $\lambda=8$.}
\begin{tabular}{cccc}
    \toprule
     & coarse & medium & fine \\
    \midrule
    $R / \Delta_{g}$ & $24$ & $32$ & $42$ \\
    grid number & $2.78 \times 10^{6}$ & $5.94 \times 10^{6}$ & $1.27 \times 10^{7}$ \\
    $C_{P}$ & $0.534$ & $0.523$ & $0.515$ \\
    $C_{T}$ & $0.792$ & $0.786$ & $0.781$ \\
    \bottomrule
\end{tabular}
\end{table}

\begin{figure}
\centering
\includegraphics[width=0.8\textwidth]{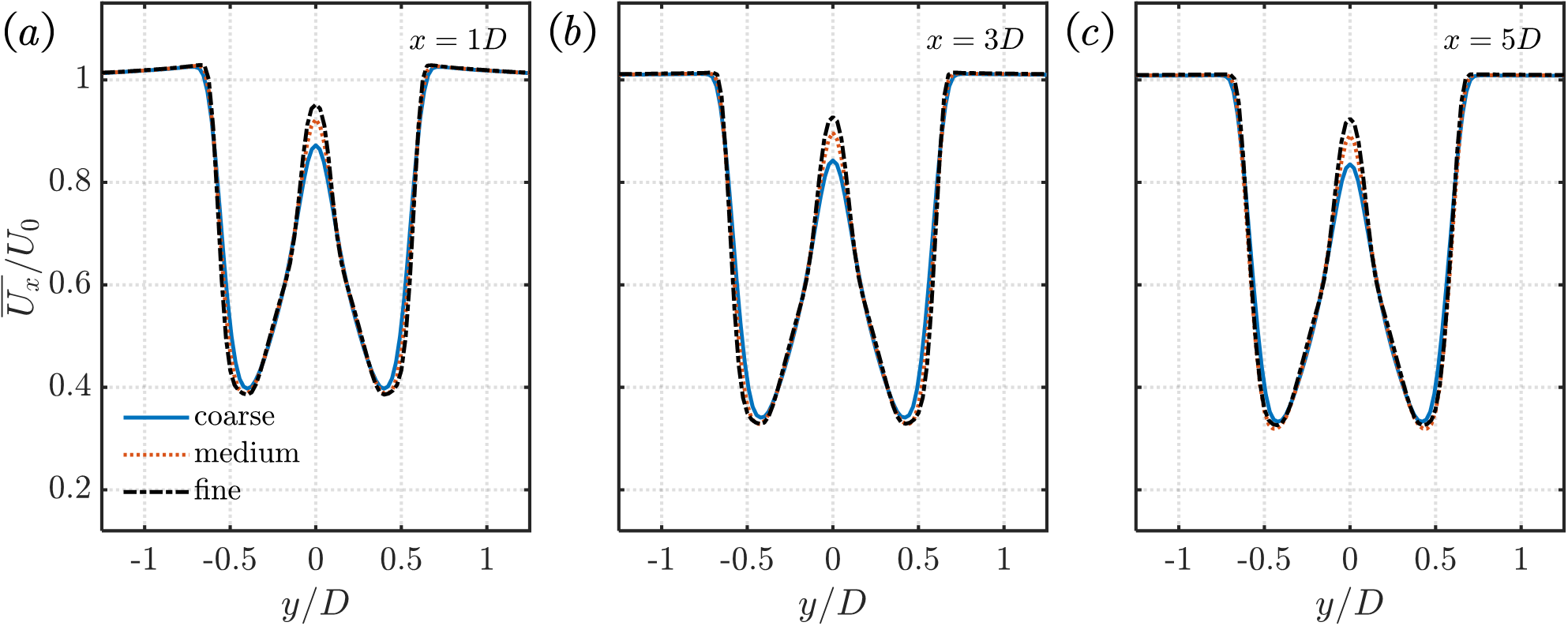}
\caption{\label{fig:validation_wake}Wakes profiles of the wake of NREL 5 MW rotor at $\lambda=8$. ($a$) $x=1D$, ($b$) $x=3D$, ($c$) $x=5D$. }
\end{figure}

\begin{figure}
\centering
\includegraphics[width=0.85\textwidth]{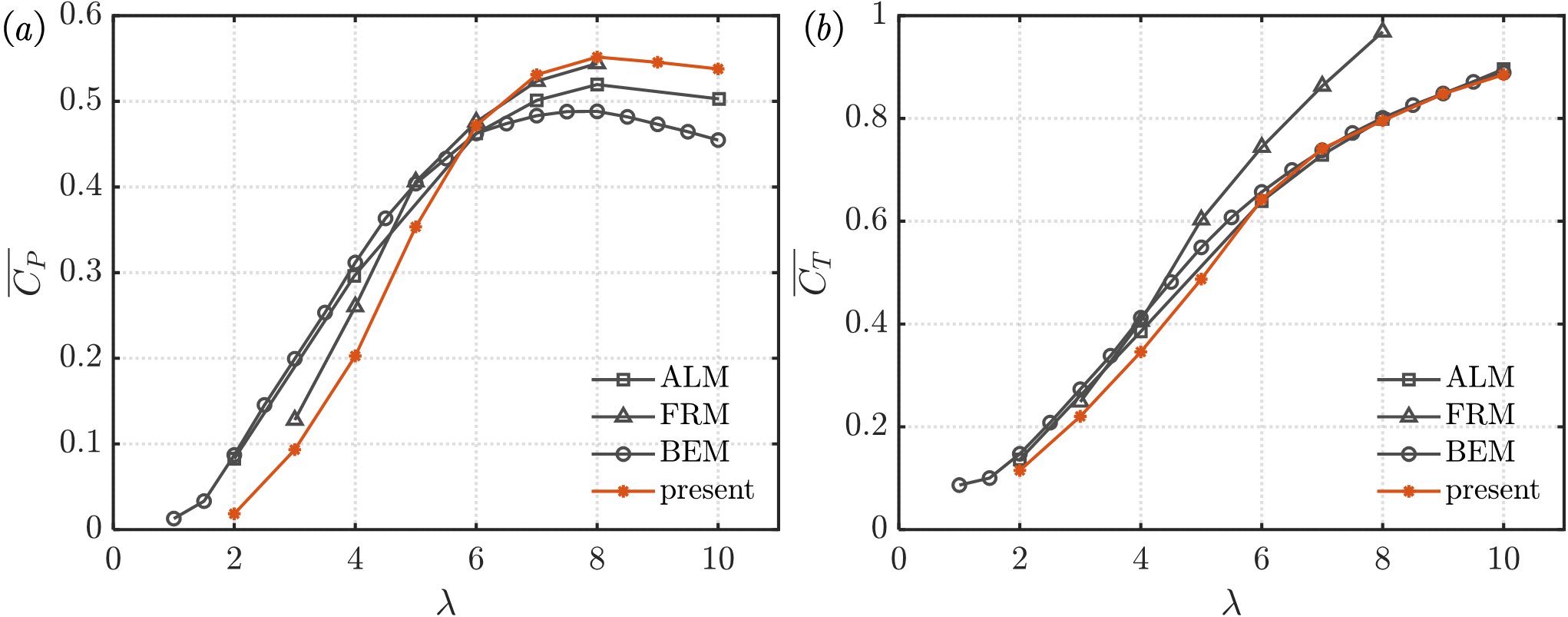}
\caption{\label{fig:validation_papers}($a$) Power coefficients and  ($b$) thrust coefficients of NREL 5 MW under different TSR.
The square line represents the LES with ALM by \citet{onel2021investigation}.
The pentagram line is using Unsteady Reynolds-Averaged Simulation (URANS) coupled with fully-resolved model (FRM) by \citet{13} and \citet{14}. 
The hollow circle line is from the blade element momentum theory in Qblade by \citet{15}. 
The red circle line is the results from the present study.}
\end{figure}

In this section, we carry out detailed mesh dependency test and compare our results with the literature.
Three different resolutions, $R / \Delta_{g} = 24,32,42$ (where $\Delta_{g}$ is the grid size around the blade, $R$ is the rotor radius) are employed to validate the grid independence.
For a single NREL 5 MW rotor simulation at tip speed ratio of $\lambda=8$, the grid numbers and results for the three resolutions are given in table \ref{tab:turbine}. 
Here, the power and thrust coefficients of the rotor are defined as 
\begin{equation}
    C_P = \frac{P}{\rho U_{\infty}^3 A/2}, \quad C_T = \frac{T}{\rho U_{\infty}^2 A/2},
    \label{equ:CPCT}
\end{equation}
where $P$ and $T$ are the power and thrust force on the rotor, and $A=\pi R^2$ is the swept rotor area. 
With increasing grid resolution, the power coefficients and thrust coefficients vary only slightly.
The wake velocity profiles at three downstream positions $x = 1D$, $3D$, $5D$ are shown in figure \ref{fig:validation_wake}. 
Except for the region near $y/D=0$ where the hub is not modeled, the three grid resolutions results in similar wake velocity distributions.
With the convergence of the wake velocity profiles in the far wake, accurate results can be expected in the two tandem turbines cases.
For the rest of the paper, the simulations are carried out using the medium grid resolution.

To further validate our numerical setup, we compute the power and thrust coefficients at different tip speed ratios for the single turbine case, and compare them with the literature in figure \ref{fig:validation_papers}.
The power coefficient of the rotor increases with tip speed ratio initially and reaches peak at $\lambda=8$, while the thrust coefficient increases monotonically with $\lambda$.
Overall, both the power and thrust coefficients predicted by the present ALM simulations agree well with the literature, showcasing the correctness of the present numerical setup.


\section{Results}
\label{sec:results}
In this section, we present the results from the ALM simulations of the two tandem turbines.
In section \ref{sec:aerodynamicPerformance}, we discuss the aerodynamic performance of the two rotors under yaw control.
The wake profiles are then shown in section \ref{sec:wakeProfiles}.
At last, the unsteady aerodynamic properties of the tandem turbines are presented in section \ref{sec:unsteady}.
In these discussions, we also incorporate the results from the low-fidelity modeling tool FLORIS \citep{FLORIS_2021} as a comparison where possible.

\subsection{Aerodynamic performance}
\label{sec:aerodynamicPerformance}

\begin{figure}
\centering
\includegraphics[width=1\textwidth]{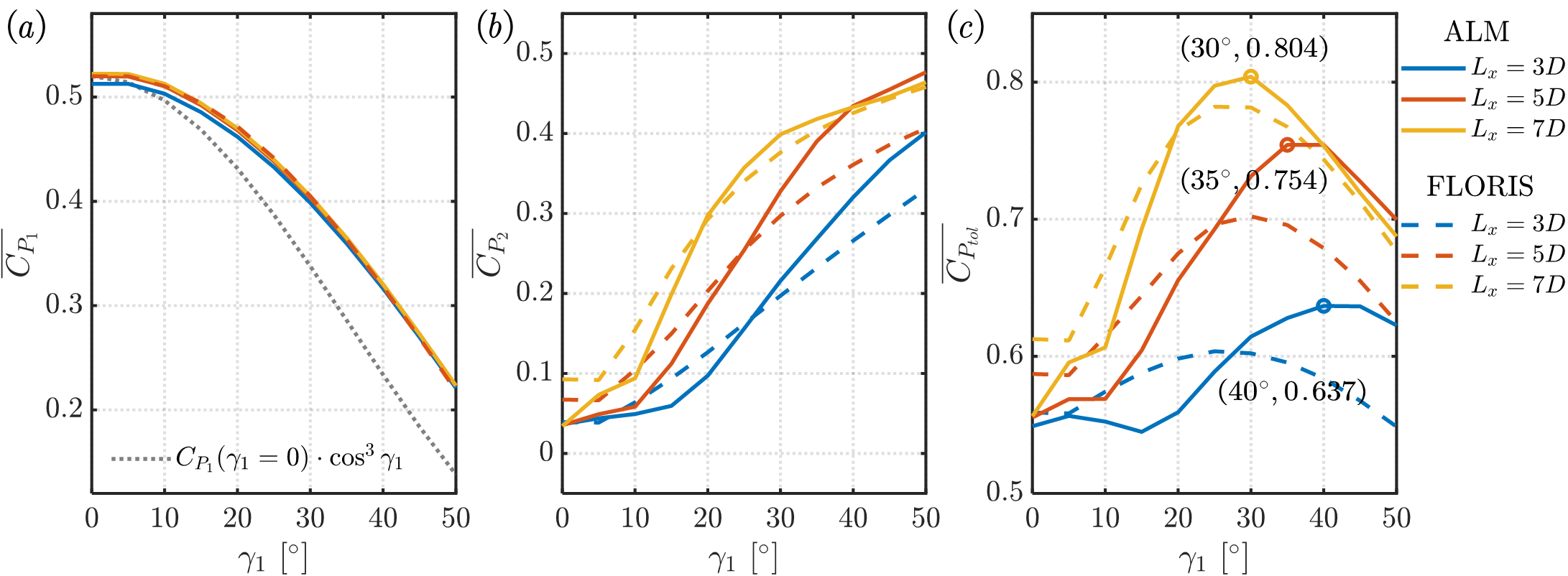}
\caption{\label{fig:coyawcp1cp2cptol}The power coefficients of ($a$) upstream rotor, ($b$) downstream rotor, ($c$) two rotor combined. The gray dashed line in ($a$) is suggested by \citet{inbook}.}
\end{figure}

\begin{figure}
\centering
\includegraphics[width=0.7\textwidth]{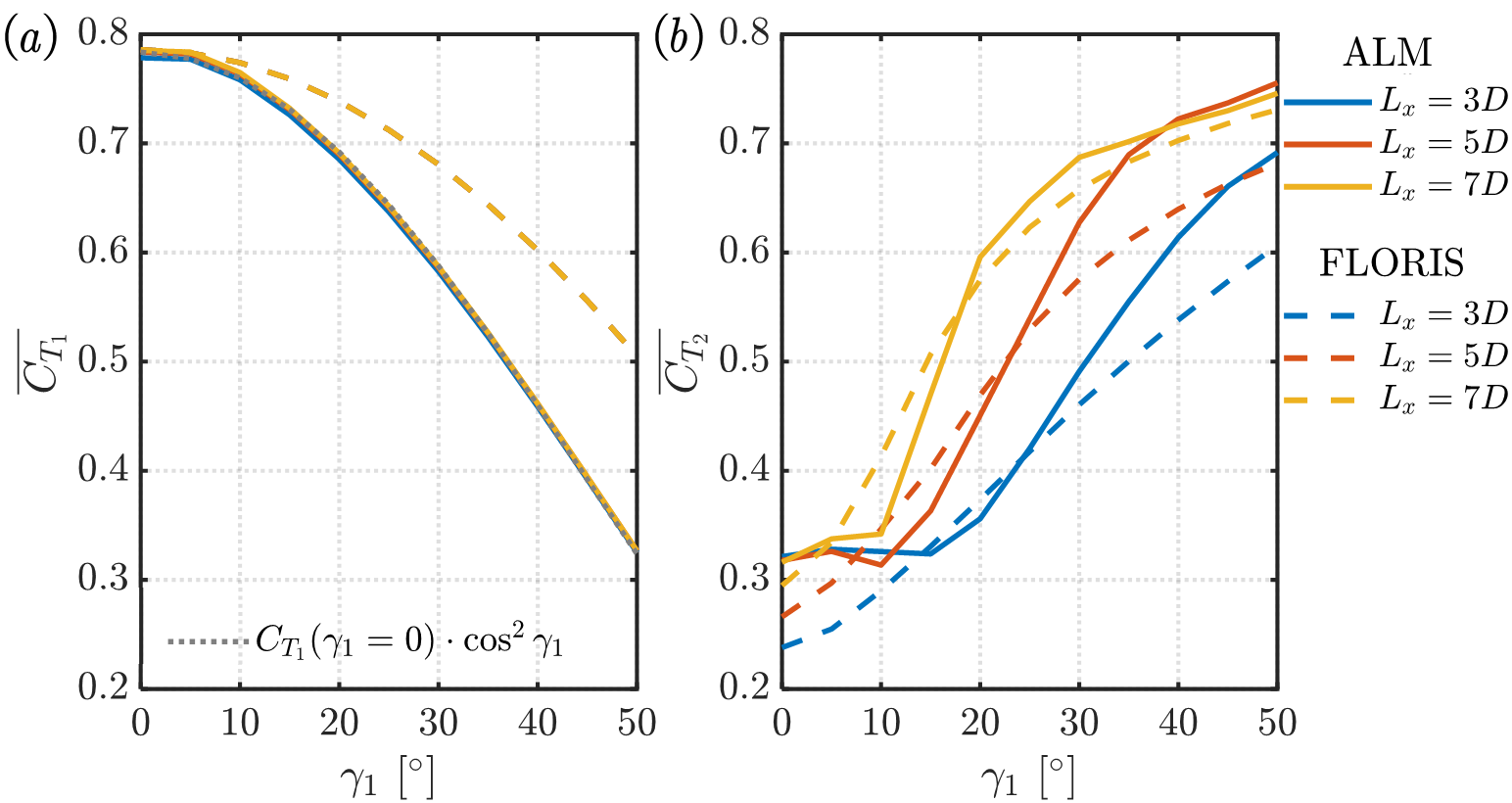}
\caption{\label{fig:coyawct}The thrust coefficients of ($a$) upstream rotor and ($b$) downstream rotor. The other two lines is covered by the yellow dashed line in ($a$).}
\end{figure}

\begin{figure}
\centering
\includegraphics[width=0.95\textwidth]{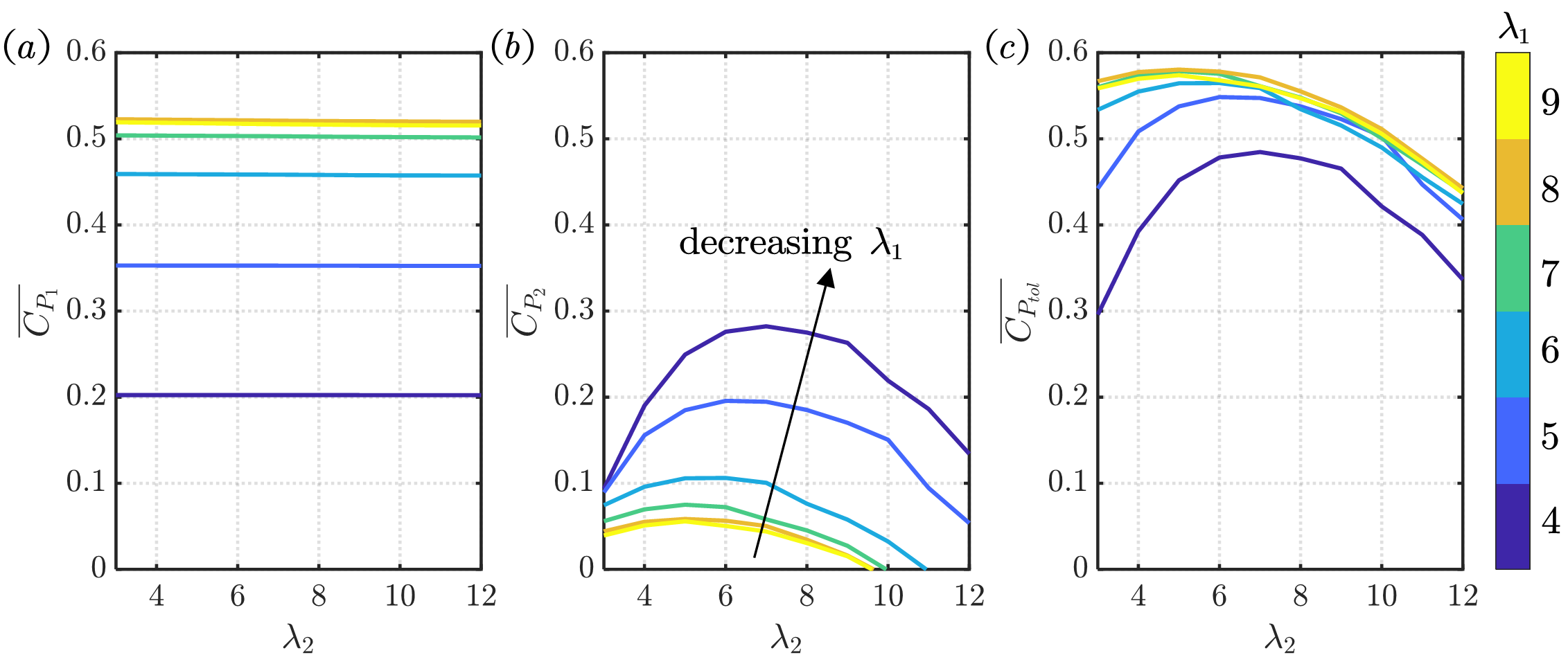}
\caption{\label{fig:TSRCps} Power coefficients in torque control. ($a$) upstream turbine, ($b$) downstream turbine, and ($c$) two turbines combined for $L_x=5D$. }
\end{figure}

The power coefficients of the two turbines under varying yaw angles of the upstream rotor ($\gamma_1$) and at different spacings ($L_x=3D$, $5D$ and $7D$) are shown in figure \ref{fig:coyawcp1cp2cptol}.
For the upstream rotor, the power coefficient decreases with $\gamma_1$, since the component of wind velocity which is normal to the rotor decreases.
Most analytical wind farm power models assume that the power of a yawed rotor follows $P(\gamma_1)=P(0)\cos^{P_p}(\gamma_1)$.
Based on the classical one-dimensional momentum theory with an incoming axial freestream wind speed of $U_{\infty}\cdot \cos(\gamma_1)$ perpendicular to the rotor, \cite{inbook} instructed that the power production of a yawed wind turbine decreases following $\cos^3(\gamma_1)$, i.e., $P_p=3$, as shown by the dotted line in figure \ref{fig:coyawcp1cp2cptol}($a$).
This is clearly not in agreement with the current ALM simulations.
In fact, the momentum theory neglects the dependence of the induction incurred by the rotor yaw.
The value of $P_p$ is reported to be widespread depending on the turbine model, typically between $1<P_p<3$ \citep{schreiber2017verification,liew2020analytical,howland2020influence}.
We observe that the relationship between $C_{P_1}$ and $\gamma_1$ calculated by ALM is close to that predicted by FLORIS, which employs a default value of $P_p=1.88$ \citep{annoni2018analysis} for the NREL 5 MW turbine.
As expected, the spacing between the two rotors has negligible effects on the power generation of the upstream one.

For the downstream rotor, the power coefficient generally increases with the yaw angle of the upstream rotor, as depicted in figure \ref{fig:coyawcp1cp2cptol}($b$).
With small spacing $L_x=3D$, the increment of power coefficients of the downstream rotor is not significant for $\gamma_1\lesssim 15^{\circ}$.
For larger yaw angles, the $C_{P_2}$-$\gamma_1$ curve exhibits almost linear growth up to $\gamma_1=50^{\circ}$.
With larger spacings $L_x=5D$ and $7D$, the downstream rotor generates more power than with $L_x=3D$.
But the growth rate of $C_{P_2}$ with respect to $\gamma_1$ gradually saturates at higher yaw angles.
This is due to the fact that the deflected upstream wake bypasses the downstream rotor, as will be discussed in section \ref{sec:wakeProfiles}.

The power coefficient of the downstream rotor predicted by the low-fidelity model FLORIS is also presented in figure \ref{fig:coyawcp1cp2cptol}($b$).
The FLORIS code calculates the aerodynamic performance by looking up the input table of the NREL 5 MW turbine \citep{jonkman2009definition}, based on the averaged velocity over the downstream rotor area \citep{annoni2018analysis}.
Since the freestream velocity is fixed at the rated wind speed $U_{\infty}=11.4$ m/s in this study, the averaged wind velocity on the downstream rotor is always smaller than $U_{\infty}$ due to wake effects.
In the below-rated operating condition region 2, the blade pitch angle is fixed at zero, and the rotor speeds increase linearly with wind speed to maintain constant optimal tip speed ratio around $\lambda=8$, which is in the same setting with ALM.
Although exact match with the ALM results is not achieved, the power coefficients of the downstream rotor calculated by FLORIS also feature an increasing trend with growing $\gamma_1$.
The positive effect of $L_x$ on the power generation of the downstream rotor is also predicted in FLORIS.

\begin{table}
\renewcommand*{\arraystretch}{1.2}
\centering
\caption{\label{tab:CpCompare}The comparison of optimal performance in yaw misalignment cases between ALM and FLORIS results. $\Delta C_{P_{tol}}= \frac{C_{P_{tol}}(\gamma_1^{opt})-C_{P_{tol}}(0^\circ)}{C_{P_{tol}}(0^\circ)}$.}
\begin{tabular}{c|ccc|ccc}
    \toprule
& \multicolumn{3}{c|}{ALM} & \multicolumn{3}{c}{FLORIS}\\
$L_x/D$ & $3$ & $5$ & $7$ & $3$ & $5$ & $7$\\
\midrule
$\gamma_1^{opt}$ & $40^\circ$ & $35^\circ$ & $30^\circ$ & $25^\circ$ & $30^\circ$ & $25^\circ$\\
$C_{P_{tol}}(\gamma_1^{opt})$ & $0.637$ & $0.754$ & $0.804$ & $0.604$ & $0.702$ & $0.782$\\
$C_{P_{tol}}(0^\circ)$ & $0.549$ & $0.555$ & $0.556$ & $0.559$ & $0.587$ & $0.613$\\
$\Delta C_{P_{tol}}$ & $16.0\%$ & $35.9\%$ & $45.0\%$ & $8.0\%$ & $19.6\%$ & $27.7\%$\\
    \bottomrule
\end{tabular}
\end{table}

The total power coefficients, $C_{P_{tol}} = C_{P_1}+C_{P_2}$, exhibit a nonmonotonic relationship with $\gamma_1$.
This is a result of the decreasing $C_{P_1}$ and increasing $C_{P_2}$, as the yaw angle of the upstream rotor increases.
For $L_x=3D$, maximum total power output is reached when the upstream rotor is yawed at $\gamma_1\approx 40^{\circ}$.
With increasing spacing between the two rotors, the optimal yaw angle decreases, and the maximum total power increases.
Compared to the unyawed cases, the maximum total power output with yaw control increases by 16.0\%, 35.9\% and 45.0\% for $L_x=3D, 5D$ and $7D$, respectively.
It is noted that with $L_x=5D$ and $7D$, the optimal yaw angle is close to the $\gamma_1$ at which the growth rate of $C_{P_2}$ saturates.
The total power curves predicted by FLORIS also exhibit bell shape, similar to the ALM results.
While the optimal yaw angle of the $L_x=3D$ case predicted by FLORIS is far from that obtained by ALM, the agreement is closer for cases with $L_x=5D$ and $7D$.

The thrust coefficients of the two rotors are presented in figure \ref{fig:coyawct}.
Although both ALM and FLORIS predict downward trend of $C_{T_1}$ with growing $\gamma_1$, the agreement between these two methods is not satisfactory compared with the power coefficients.
The ALM results show that the thrust coefficients of the yawed upstream rotor follow $C_{T_1}(\gamma_1)=C_{T_1}(0)\cdot\cos^2(\gamma_1)$, while in FLORIS the scaling factor is set as $\cos(\gamma_1)$ by default.
The thrust coefficients of the downstream rotor increase with the upstream yaw angle in a similar fashion with the power coefficients.

Let us compare the effectiveness of yaw control against induction control.
Here, the induction control is realized by changing the rotating speed of the two rotors, while the yaw angle and blade pitch angle are fixed at zero.
The spacing between the two rotors is fixed at $L_x=5D$.
The tip speed ratio of the upstream rotor ranges from $\lambda_1=4$ -- 9.
For each $\lambda_1$, the tip speed ratio of the downstream rotor $\lambda_2$ is also varied to locate the optimal operating condition that results in maximum power output.
The power coefficients of the two rotors are shown in figure \ref{fig:TSRCps}($a,b$).
As the upstream rotor is derated from $\lambda_1=9$ to 4, the optimal $\lambda_2$ shifts to higher values, and the power generation of the downstream rotor is enhanced.
Nevertheless, the maximum power coefficients of the two rotors combined is only 0.58, which is much lower than that could be achieved with yaw control.
This comparison suggests greater effectiveness of yaw control over induction control, as also noted in other studies \citep{nash2021wind,houck2022review,li2022study}.

We note in passing that additional simulations with negative yaw angles have also been carried out, arriving at the conclusion that both positive and negative yaw angles yield the same power production.
This is in contradiction to the findings that direction of yaw angle has noticeable influence on the total power \citep{schottler2016wind,fleming2018,bartl2018wind_b}.
\citet{archer2019wake} hypothesized that the difference of power production between positive and negative yaw angles is related to the Coriolis effect, and recommended that only positive yaw misalignment angles should be considered for wake steering purposes in the northern hemisphere.
Another explanation put forward by 
\citet{fleming2018} suggests that the difference is due to the ground effects and wind shear. 
As will be discussed in \S \ref{sec:wakeProfiles}, a yawed turbine creates a highly three-dimensional wake featuring a pair of counter-rotating vortices at the top and bottom of the rotor, respectively.
When a turbine is positively yawed, the top vortex rotates the same direction as the wake itself (opposite the rotor rotation), which strengthens that vortex.
When negatively yawed, the lower vortex is enhanced in the same way, but it also experiences lower wind speeds and ground shear.
However when the turbine is positively yawed, the top vortex is in higher wind speeds and unencumbered by the ground, which allows it to have a greater effect on the shape of the wake.
Since neither the Coriolis force and ground effects are considered in this study, it is reasonable that the power generated by the two rotors is insensitive to the sign of yaw angle.

\subsection{Wake profiles}
\label{sec:wakeProfiles}

\begin{figure}
\centering
\includegraphics[width=0.55\textwidth]{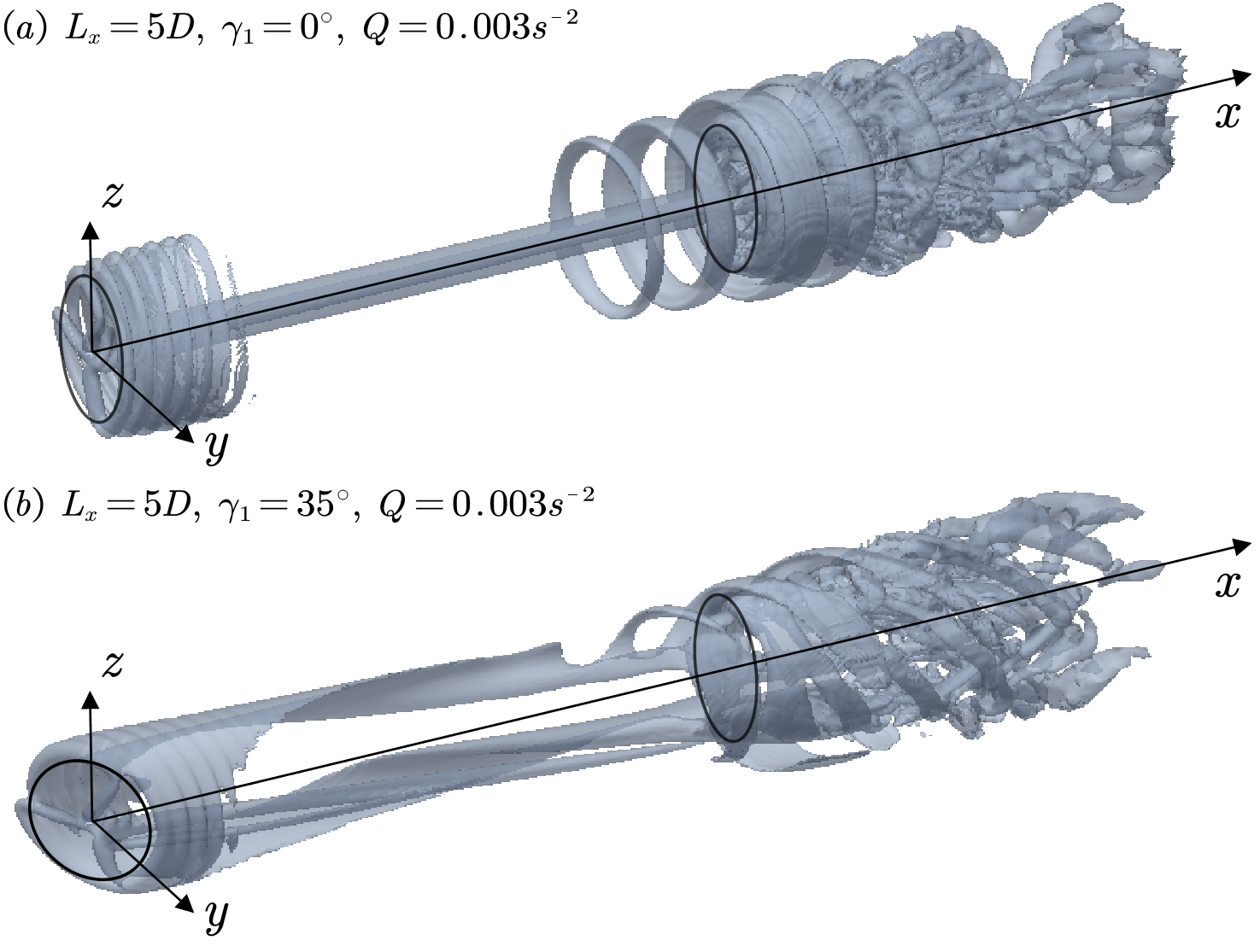}
\caption{\label{fig:q}Instantaneous vortical structures visualized by iso-surfaces of the $Q$-criterion for ($a$) $L_x=5D,\gamma_1=0^\circ$ and ($b$) $L_x=5D, \gamma_1=35^\circ$ cases. The black circles denote the position of two rotors.}
\end{figure}

\begin{figure}
\centering
\includegraphics[width=0.9\textwidth]{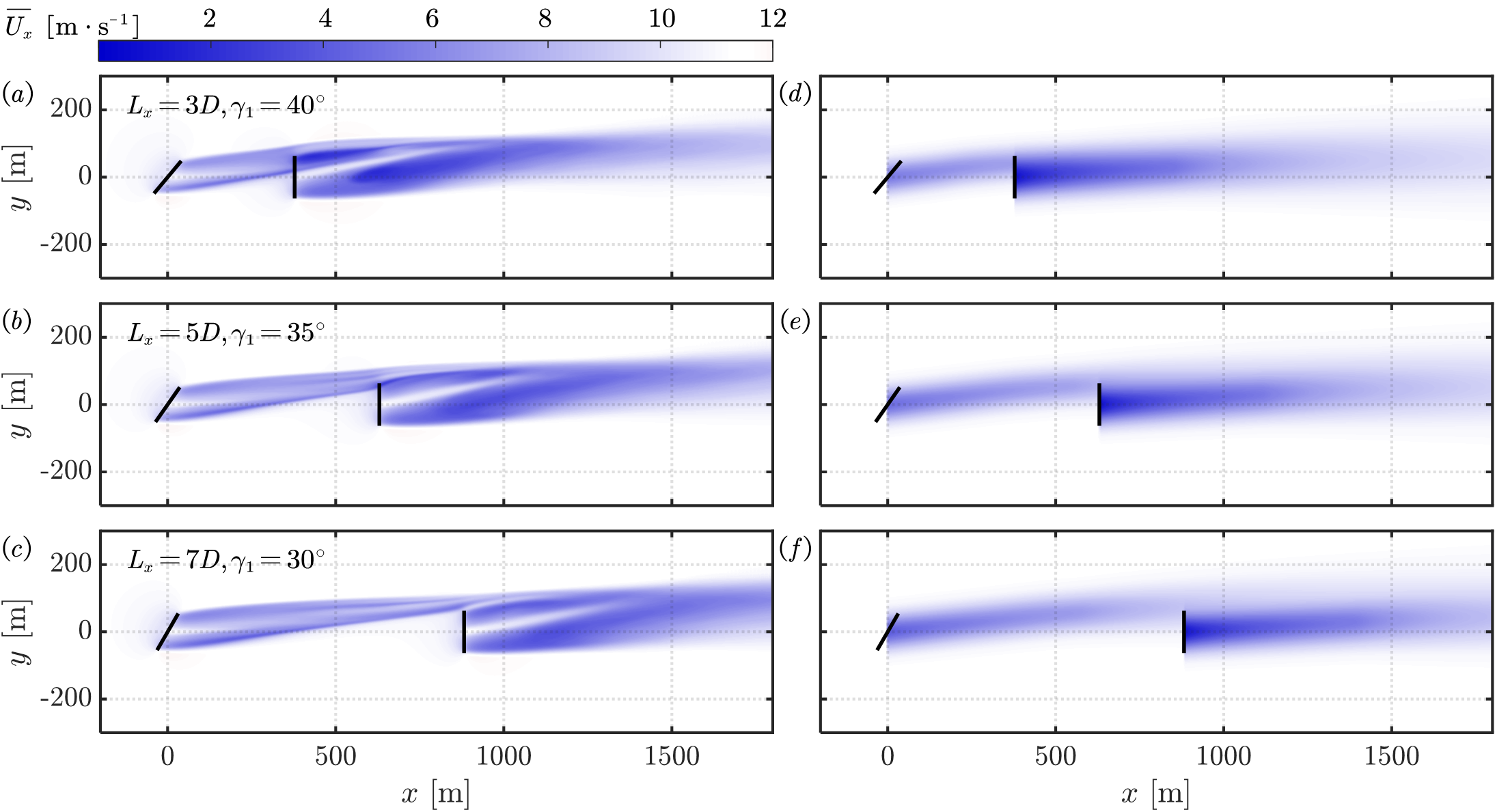}
\caption{\label{fig:z0WakeCompile}Time-averaged streamwise velocity fields calculated by ALM ($a,b,c$) and FLORIS ($d,e,f$) on $z=0$ plane. Shown are the cases of optimal yaw angle for $L_x=3D$, $5D$ and $7D$, respectively.}
\end{figure}

\begin{figure}
\centering
\includegraphics[width=0.9\textwidth]{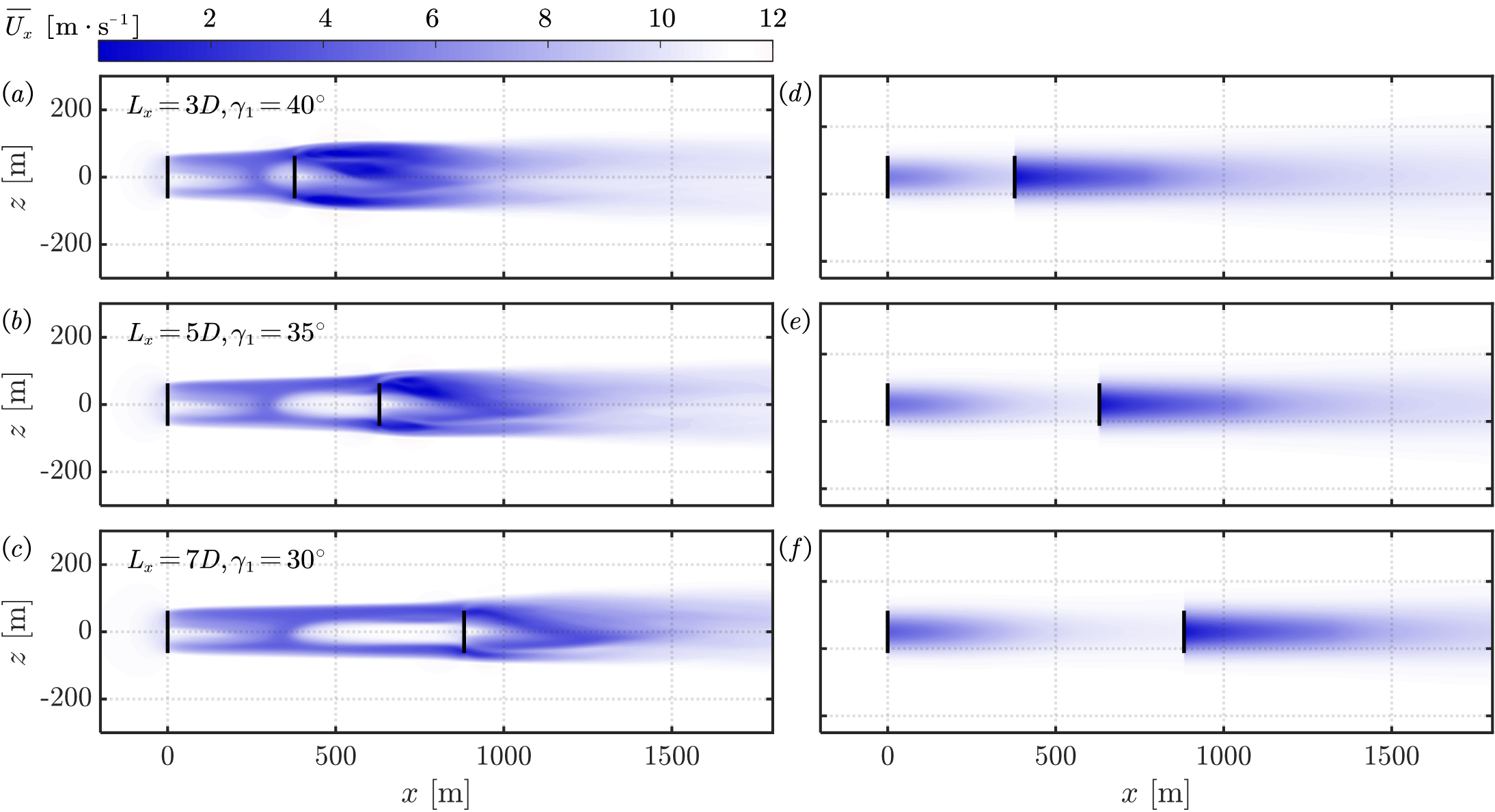}
\caption{\label{fig:StreamUxOnXZFloris}Time-averaged streamwise velocity fields calculated by ALM ($a,b,c$) and FLORIS ($d,e,f$) on $y=0$ plane. Shown are the cases of optimal yaw angle for $L_x=3D$, $5D$ and $7D$, respectively.}
\end{figure}

\begin{figure}
\centering
\includegraphics[width=0.9\textwidth]{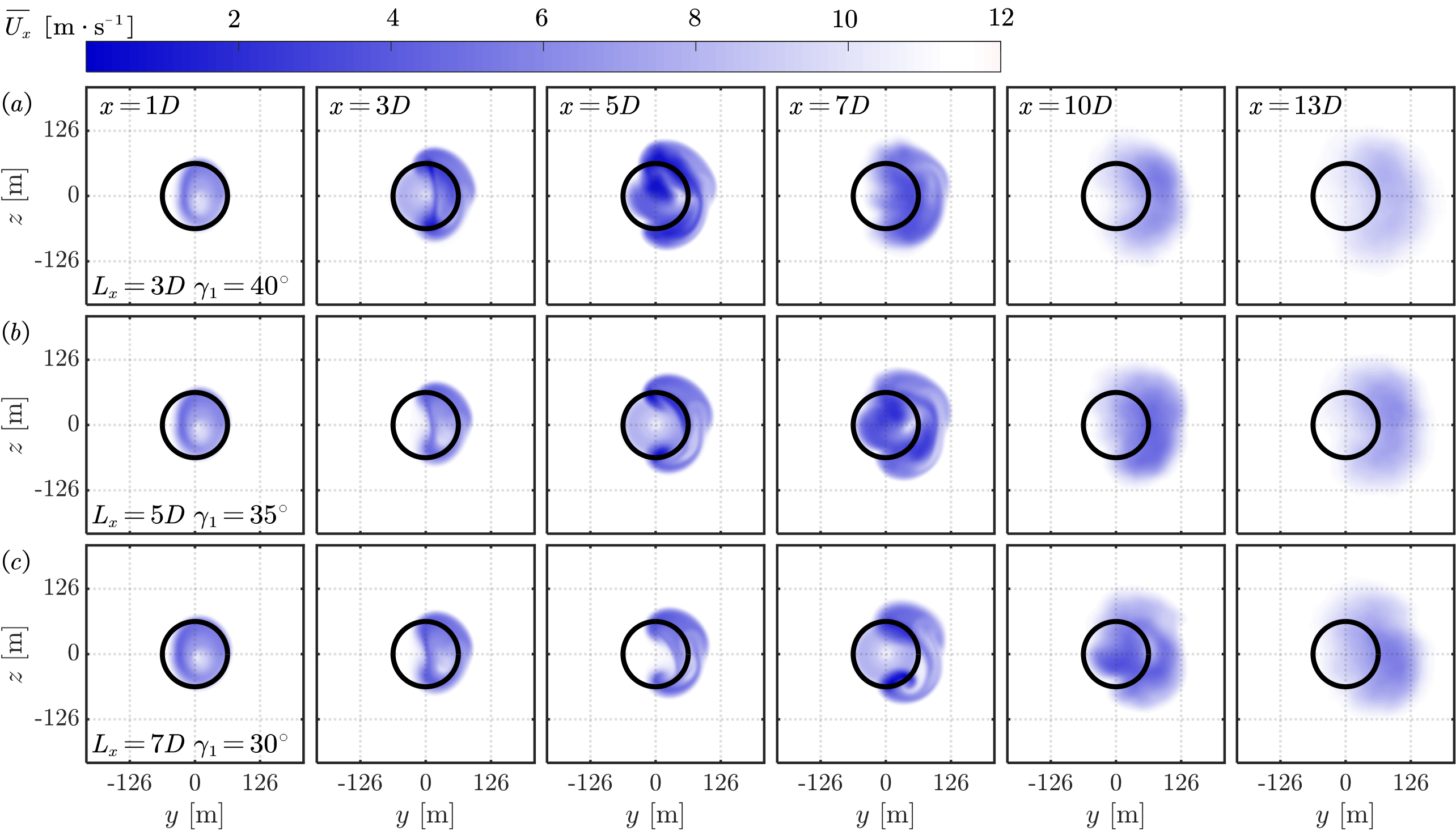}
\caption{\label{fig:yzwakeCompare}Time-averaged streamwise velocity fields calculated by ALM in optimal cases of two tandem turbines. ($a$) $L_x=3D$, $\gamma_1=40^\circ$, ($b$) $L_x=5D$, $\gamma_1=35^\circ$, ($c$) $L_x=7D$, $\gamma_1=30^\circ$. The cloud maps are positioned at $x=1D$, $3D$, $5D$, $7D$, $10D$, $13D$. The black circle denotes the position of an non-yawed rotor.}
\end{figure}

\begin{figure}[!h]
\centering
\includegraphics[width=0.9\textwidth]{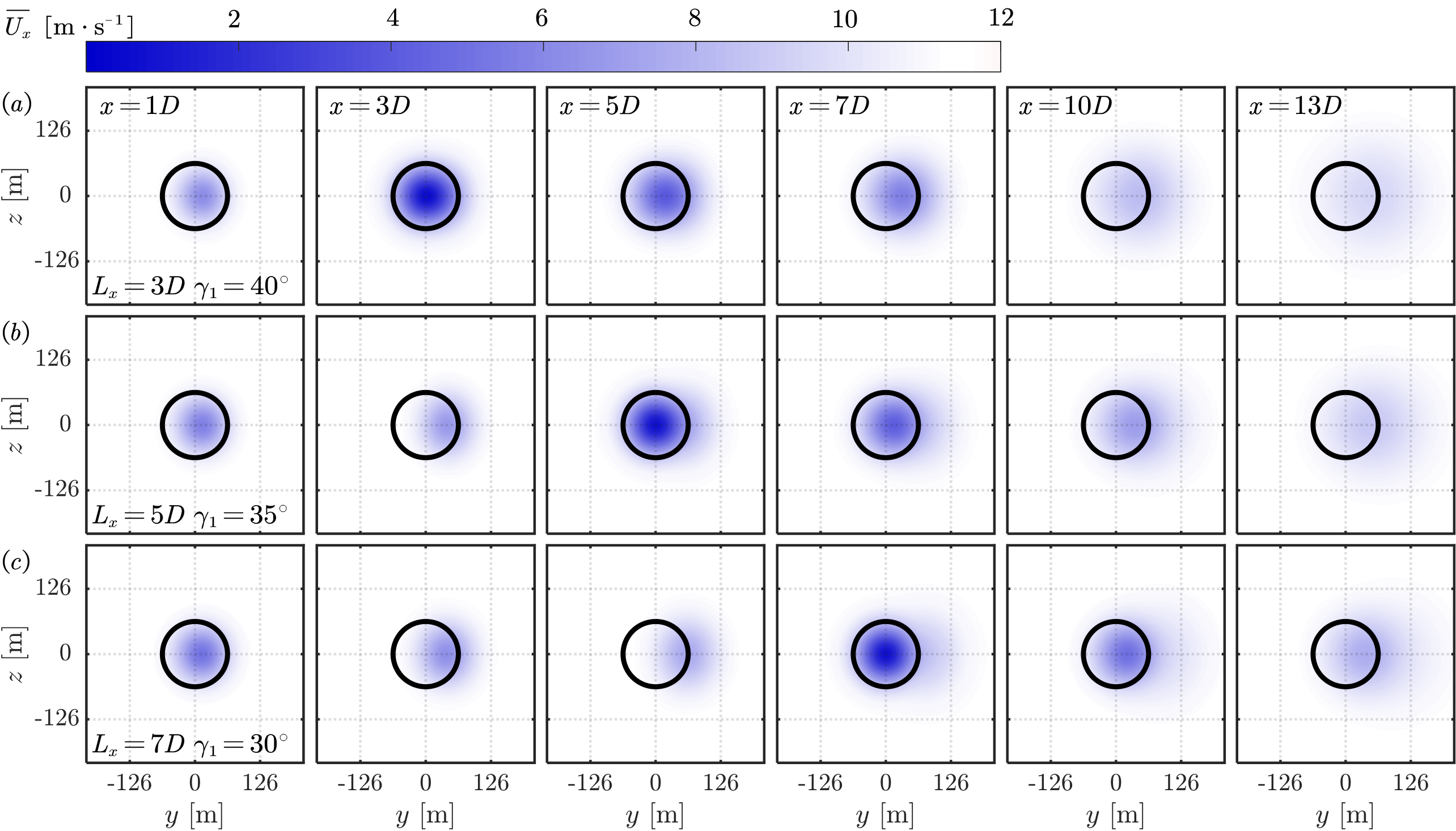}
\caption{\label{fig:yzwakeCompare_floris}Time-averaged streamwise velocity fields calculated by FLORIS in optimal cases of two tandem turbines. ($a$) $L_x=3D$, $\gamma_1=40^\circ$, ($b$) $L_x=5D$, $\gamma_1=35^\circ$, ($c$) $L_x=7D$, $\gamma_1=30^\circ$. The cloud maps are positioned at $x=1D$, $3D$, $5D$, $7D$, $10D$, $13D$.The black circle denotes the position of an non-yawed rotor.}
\end{figure}

We analyze the wake profiles with the aim to shed light upon the aerodynamic performance described above. 
Figure \ref{fig:q} shows instantaneous vortical structures visualized by isosurfaces of $Q$ (second invariant of velocity gradient tensor) for the cases with $\gamma_1=0^{\circ}$ and $35^{\circ}$ at $L_x=5D$.
For the unyawed case, the near wake of the upstream rotor is dominated by helical tip vortices shed from the blades.
As the upstream wake impinges on the downstream rotor, torus-like vortices form around the rotor, and then breaks down in the far wake.
On the other hand, the wake of the yawed upstream rotor features a pair of counter-rotating vortices that trails into the far wake.
These streamwise vortices interact with a part of the vortical structures generated at the outskirt of the downstream rotor, while the rest part (that is less affected by the deflected upstream wake) still shed helical tip vortices.
As a result, the wake of the dowmstrean rotor appear less axis-symmetric compared to the unyawed case.

The time-averaged streamwise velocity fields of the optimal cases for $L_x=3D$, $5D$ and $7D$ are shown in figures \ref{fig:z0WakeCompile} and \ref{fig:StreamUxOnXZFloris} on $x$-$y$ planes and $x$-$z$ planes, respectively.
The introduction of yaw on the upstream rotor deflects its wake away from the centerline on the $x$-$y$ planes, as is clear from both the ALM and FLORIS calculations.
For $L_x=3D$, the optimal yaw angle is achieved at $\gamma_1=40^{\circ}$. 
Under this condition, a significant portion of the upstream wake still impinges on the downstream rotor.
Although it is possible to increase the yaw angle of upstream rotor to further steer its wake, the power gain of the downstream rotor can no longer compensate the loss in the upstream one.
With $L_x=5D$ and $7D$, the deflected wakes almost bypass the entire downstream rotor at the optimal yaw angles.
With further increase in $\gamma_1$, the gain of power in downstream rotor becomes less significant, as evidenced by the saturated growth rate of $C_{P_2}$ at high yaw angles shown in figure \ref{fig:coyawcp1cp2cptol}($b$).
On the $x$-$z$ planes, the wakes of both turbines is not deflected as shown in figure \ref{fig:StreamUxOnXZFloris}.
The velocity deficit predicted by ALM exhibits hollow area downstream of the yawed turbine, which is associated with the thinning of the wake shown on the $x$-$y$ plane. 
This feature is not predicted by FLORIS, which assumes self-similarity in the wake model.


One of the key differences between the flow fields predicted by ALM and FLORIS is that in the former, the upstream wake appears to become increasingly narrow along the streamwise direction, while for the latter the wake is more dispersed. 
To understand this difference, we show the time-averaged $\overline{U_x}$ fields on slices cut at different streamwise locations in figures \ref{fig:yzwakeCompare} and \ref{fig:yzwakeCompare_floris} for ALM and FLORIS, respectively.
With increasing downstream distance, the wake deficit of the yawed turbine not only deflects in the $y$ direction, but also curls into the kidney-like shape, particularly for cases with large yaw angles.
This type of wake is characterized by a pair of counter-rotating vortices as shown in figure \ref{fig:q}($b$), and has been discussed extensively in literature \citep{medici2006measurements,howland2016wake,bartl2018wind,bastankhah2016experimental,kleusberg2020parametric}.
It is thus clear that the narrow wake observed on the $x$-$y$ slices in figure \ref{fig:z0WakeCompile} corresponds to the thin connecting part of the two counter-rotating vortices.
The wakes calculated by FLORIS use the Gauss-curl hybrid (GCH) model, which assumes self-similarity for fast computation.
As a result, the velocity deficit remains isotropic in shape while shifting in the direction of yaw, and no counter-rotating vortex pair is reproduced.

In the immediate wake of the downstream rotor, an isotropic velocity deficit is formed, and is engulfed by the kidney-like wake incurred by the yawed upstream rotor.
Further downstream, the combined wake gradually becomes diffused as the velocity recovers.
Even though the downstream rotor is aligned perpendicular with the freestream, its wake also exhibits certain degree of deflection due to the yawed incoming flow.
This phenomenon is referred to as ``secondary steering'' \citep{fleming2018}, and is important for unraveling the full potential of yaw control \citep{rak2022impact}.
As shown in figure \ref{fig:yzwakeCompare_floris}, the secondary steering phenomenon is also captured by FLORIS, although the detailed wake shape differs from that predicted by ALM.
It is noted that the Gauss-curl hybrid (GCH) wake model \citep{king2021control} employed here is tailored for reproducing secondary steering.
This phenomenon can not be captured using conventional Jensen or Gaussian wake models \citep{fleming2018}. 

\subsection{Unsteady aerodynamics}
\label{sec:unsteady}

\begin{figure}
\centering
\includegraphics[width=0.8\textwidth]{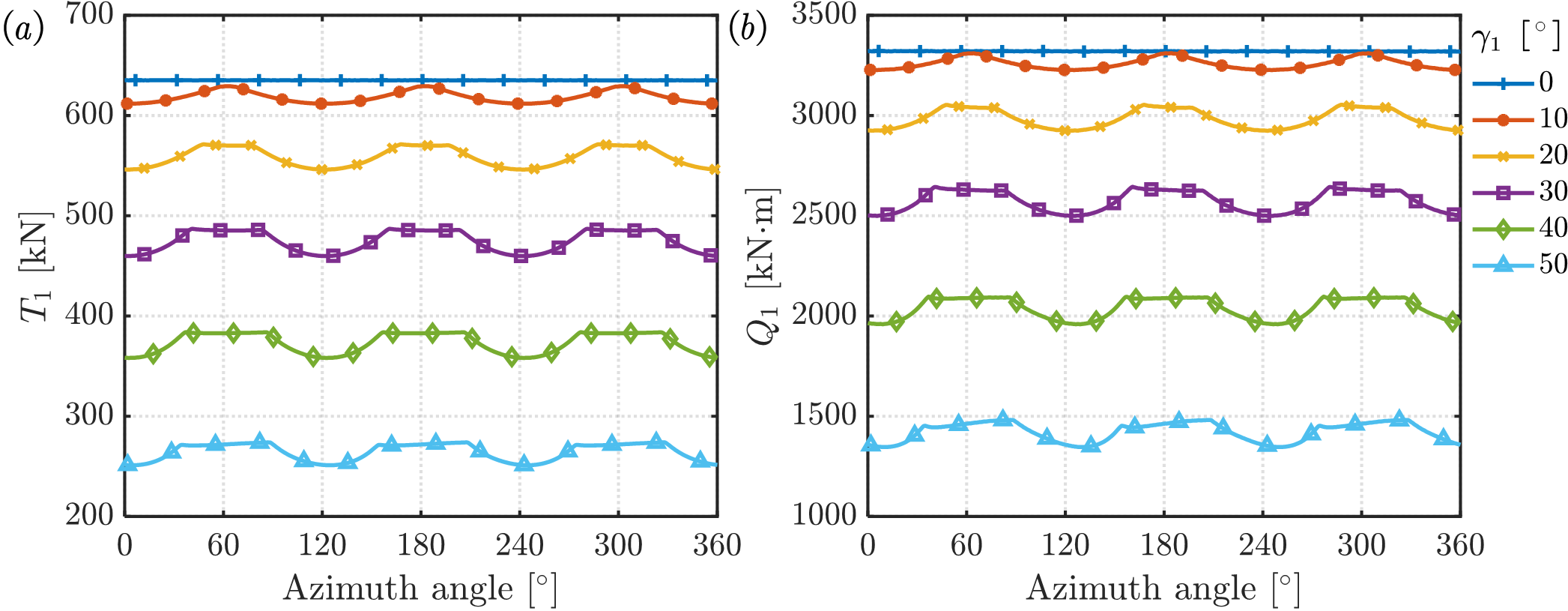}
\caption{\label{fig:360torqueThrustWT1}Variations of ($a$) thrust and ($b$) torque of upstream rotor during one rotation period under different $\gamma_1$. The shown case is with $L_x=5D$.}
\end{figure}

\begin{figure}
\centering
\includegraphics[width=0.4\textwidth]{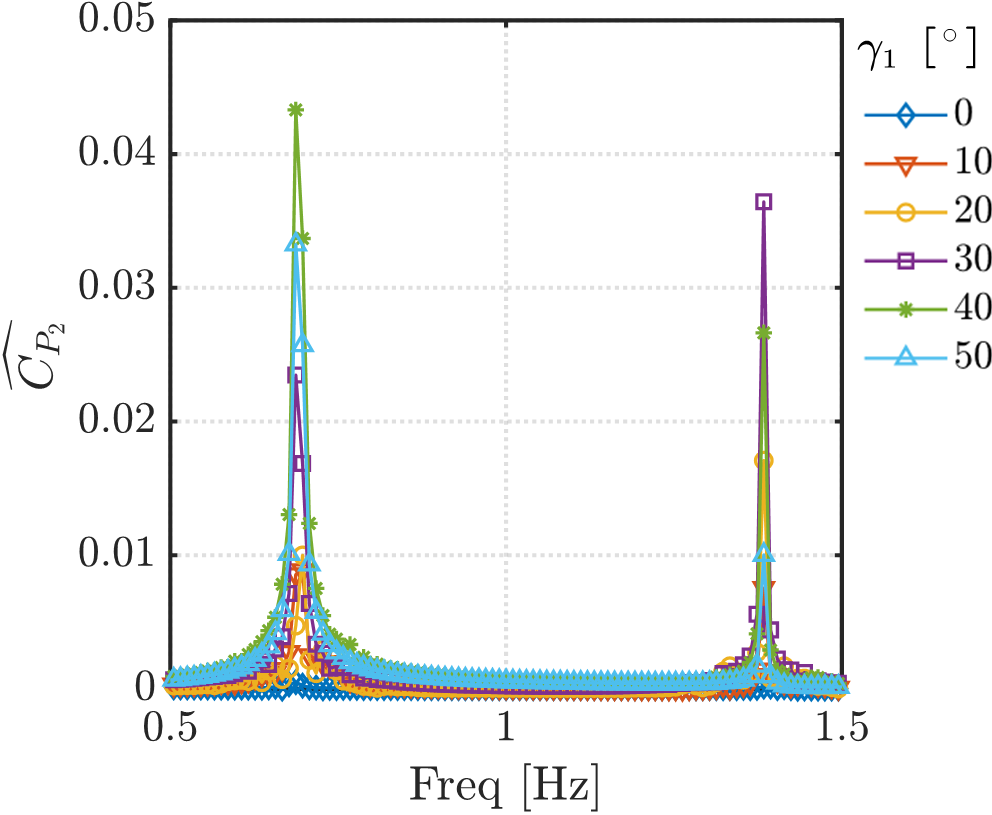}
\caption{\label{fig:freq}The power spectra of $C_{P_2}$ at different $\gamma_1$, $\gamma_1=0^\circ,25^\circ,35^\circ,40^\circ,50^\circ$. $L_x=5D$}
\end{figure}

Even in steady wind considered herein, both the yawed upstream rotor and unyawed downstream rotor experience unsteady loading, which negatively affect the power quality and fatigue life of the turbines.
For the upstream rotor, the angle of attack on each blade is continuously changing as it rotates in the yawed condition, resulting in fluctuating thrust and torque as shown in figure \ref{fig:360torqueThrustWT1}.
For the downstream rotor, the thrust and torque also exhibit periodic variation, and the fluctuations in these aerodynamic quantities are stronger than those for the upstream rotor.
Each rotation period is associated with three waves in the aerodynamic loading.
As shown in the spectrum of the power coefficients $C_{P_2}$ in figure \ref{fig:freq}, the unsteady aerodynamic performance of the downstream rotor, regardless of the yaw angle, is dominated by a peak frequency of $f_{C_P}=0.69$ Hz with a superharmonic at $2f_{C_P} = 1.38$ Hz.
This corresponds to the tip speed ratio of $\lambda=8$, which translates to a rotational frequency of $f_0=\lambda U_{\infty}/(2\pi R)=0.23$ Hz.
Since the considered turbine is three-bladed, the dominant frequency in the aerodynamic loading of the rotor is related to the rotational frequency by $f_{C_P}=3f_0$.

\begin{figure}[!h]
\centering
\includegraphics[width=0.8\textwidth]{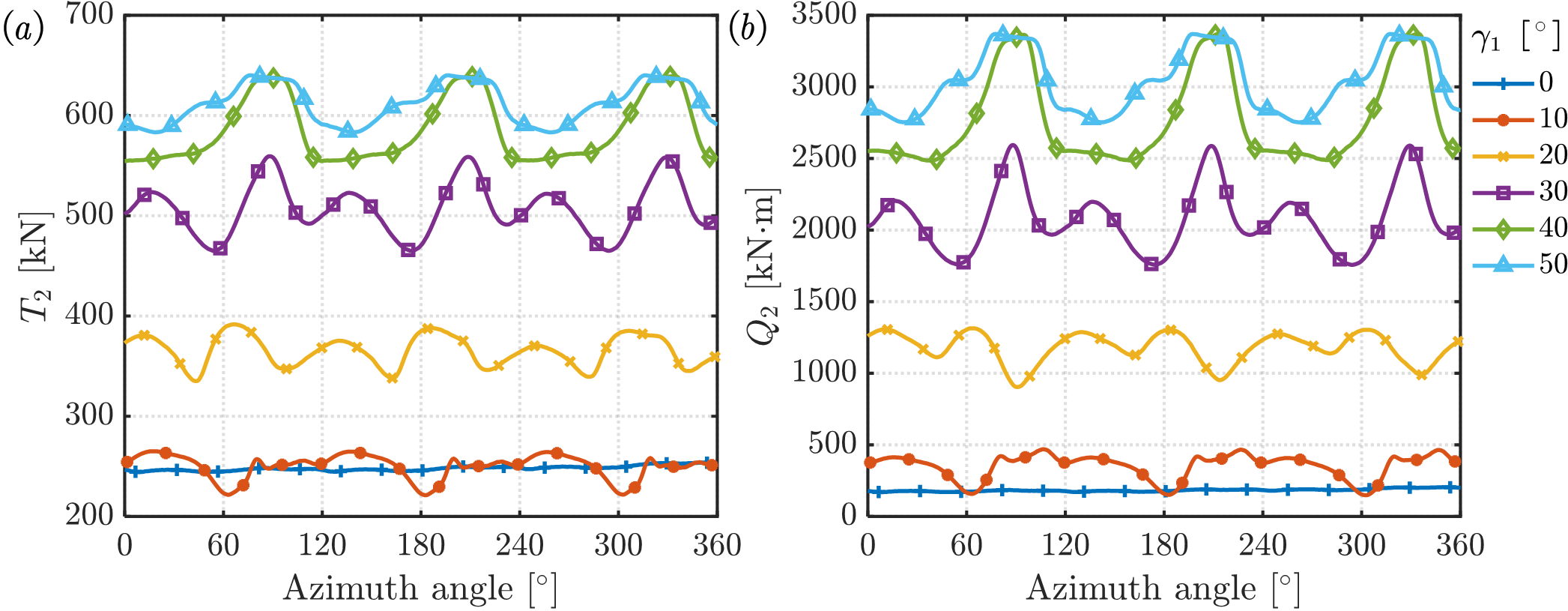}
\caption{\label{fig:360torqueThrust}Variations of ($a$) thrust and ($b$) torque of downstream rotor during one rotation period under different $\gamma_1$. The shown case is with $L_x=5D$.}
\end{figure}

\begin{figure}[!h]
\centering
\includegraphics[width=1\textwidth]{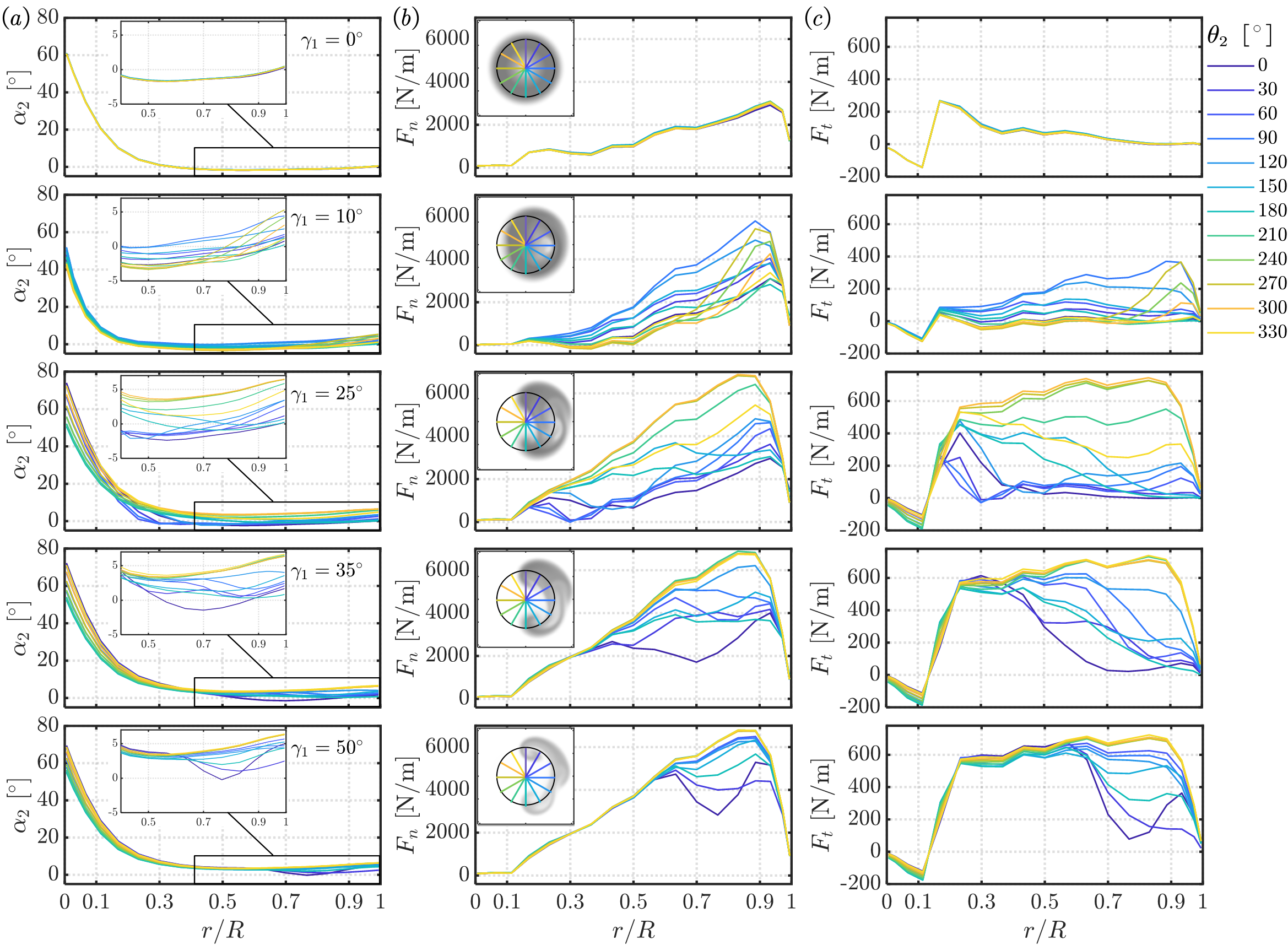}
\caption{\label{fig:spanwise}Variations of ($a$) local attack angle $\alpha_2$, ($b$) axial and ($c$) tangential force per unit span along the blade 1 of WT2 during the rotation period under different yaw angle. Shown is the case with $L_x = 5D$. $\theta_2$ is the rotational angle of blade 1. In ($a$), the insets show the zoomed view of the angle of attack for $r/R=0.4-1$. In ($b$), the insets show the wake deficit (gray) at $x=4D$ and blade positions at different azimuthal angles.}
\end{figure}

\begin{figure}
\centering
\includegraphics[width=0.8\textwidth]{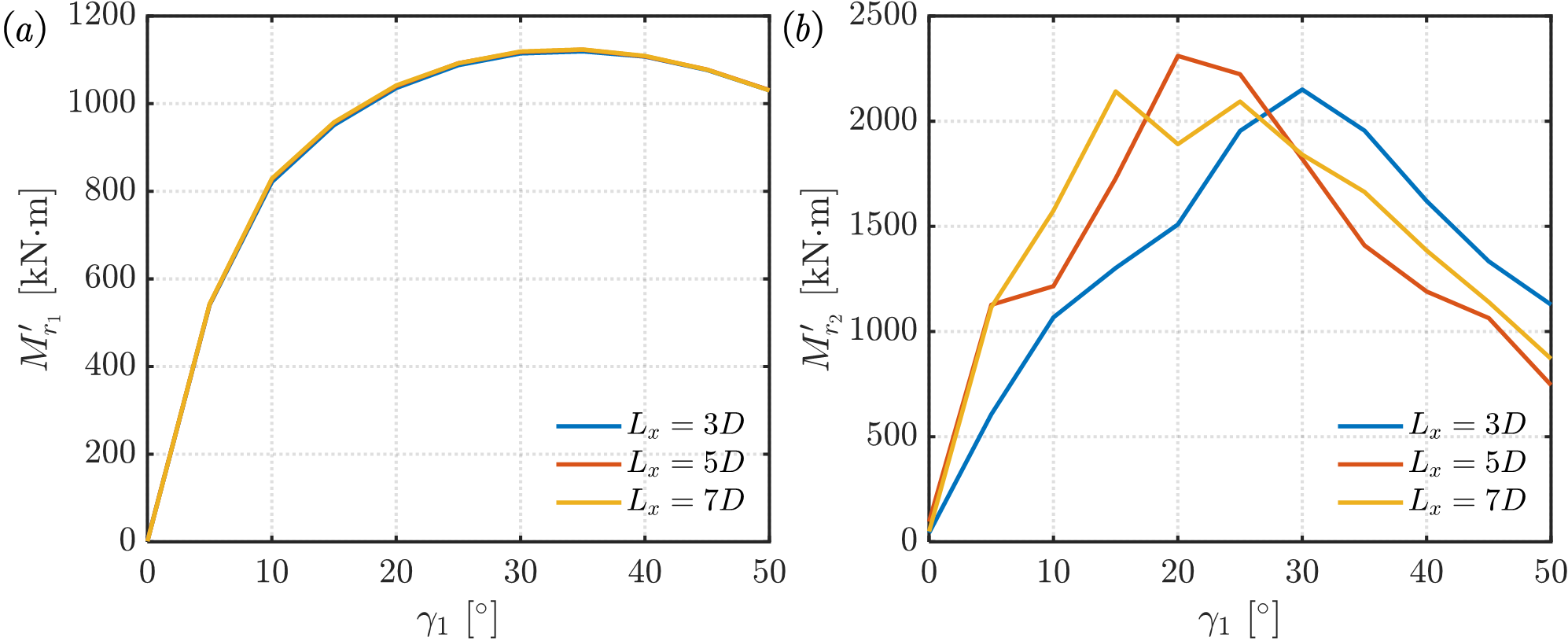}
\caption{\label{fig:momenttwo} The standard deviation of bending moment at the blade root of the upstream rotor ($a$) and downstream rotor ($b$). }
\end{figure}

The highly unsteady aerodynamic response of the downstream rotor is a result of the non-uniform wake profiles incurred by the yawed upstream rotor.
Figure \ref{fig:spanwise} shows the variations of the angle of attack, axial and tangential forces along the blade span during one rotation period of the rotor.
For the non-yawed case, since the wake profile on the $y$-$z$ plane is isotropic, the angle of attack along the blade span remains fixed over time, leading to constant aerodynamic loading.
For the yawed cases, the wake of the upstream rotor is directed away from the centerline, and deforms into a kidney-like shape, as described in \S \ref{sec:wakeProfiles}.
As the blades revolves into and out of the wake deficit, the angle of attack on the blade section changes, leading to the variations in the sectional forces. 
It is observed that with medium yaw angles, the variations in the sectional forces commence near the blade root.
For high yaw angles, since the upstream wake is almost completely deflected from the downstream rotor, only the tip region is affected by this unsteady effect, while the loading on the rest of the span remains constant.

We further present the fluctuating bending moments at the blade root of the rotors in figure \ref{fig:momenttwo}.
The bending moment is calculated as $M_r=\sqrt{M_n^2+M_t^2}$, where $M_{n,t}=\int_0^R F_{n,t}r\mathrm{d}r$ is the moment associated with the normal and tangential forces, $r$ denotes the spatial coordinate along the span.
The fluctuating bending moments of both rotors exhibit nonmonotonic relationship with $\gamma_1$.
For the upstream rotor, $M_{r_1}^{\prime}$ increases with yaw angle initially but gradually saturates at higher $\gamma_1$. 
The fluctuating bending moment of the downstream rotor is significantly higher than that of the upstream one.
While the maximum $M_{r_2}^{\prime}$ is similar among cases with different streamwise spacing, the yaw angle at which the maximum value is achieved shifts to lower value with increasing spacing.
By comparing with figure \ref{fig:coyawcp1cp2cptol}, it is noticed that the peak in $M_{r_2}^{\prime}$ occurs at smaller $\gamma_1$ than that for the combined power of the two rotors.
This is expected since the maximum total power generation is reached when the deflected upstream wake bypasses the whole downstream rotor area, while the maximum fluctuating loads occur when the upstream wake covers approximately half of the downstream rotor, which is achieved with smaller yaw angle.
The analysis of the unsteady aerodynamic performance of the two rotors serves as a precursor for assessing the fatigue loads of wind turbine blades, which is critical
for the lifespan of turbines and their maintenance cost.

\section{Conclusion}
\label{sec:conclusions}
This paper presented extensive numerical simulations to characterize the yaw control effects on the aerodynamics of two tandem turbines in uniform inflow condition. 
The simulations are performed using the mid-fidelity actuator line model, with turbulence closure by large eddy simulation.
The results from the low-fidelity modeling tool FLORIS are also included for comparison.

With increasing yaw angle, the power coefficient of the upstream rotor decreases following the $\cos^{1.88}(\gamma_1)$ relationship, and that of the downstream rotor increases more significantly, resulting in higher combined power generation.
For different spacing between the two rotors $L_x=3D, 5D$ and $7D$, the maximum total powers increase by 16.0\%, 35.9\% and 45.0\%, compared with the cases without yaw control.
The optimal yaw angle at which the maximum power is achieved occurs when the upstream wake is deflected away from the downstream rotor.
The wake of the yawed rotor is highly three-dimensional, and is featured by a pair of counter-rotating vortices resembling a kidney shape.
Although the wake shapes predicted by the low-fidelity tool FLORIS do not reveal such three dimensionality, the secondary steering phenomenon, where the wake of the downstream rotor also exhibit deflection, are captured in both models.

The use of ALM also reveals unsteady aerodynamic characteristics that can not be captured in lower-fidelity models.
Yawing the upstream rotor introduces time-varying angle of attack on the rotor blades, giving rise to the unsteady aerodynamic performance of the turbine.
For the downstream rotor, as the blades revolves into and out of the redirected upstream velocity deficit, the blades experience fluctuating aerodynamic loads with a dominant frequency dictated by the rotational speed of the rotors.
The fluctuating bending moment at the blade root of the downstream rotor is significantly higher than that of the upstream one, raising concerns of structural fatigue damage associated with yaw control.

To sum up, this paper has presented aerodynamic performance, wake profiles, and unsteady characteristics of two tandem turbines under yaw control.
The fundamental insights obtained here improves the understanding of the aerodynamics of the yaw misalignment effects on the aerodynamics of two tandem turbines, and can aid the design of collective yaw control strategies of large wind farms.

\section*{Acknowledgments}
KZ, ZLH and DZ acknowledge financial support from the Innovation Program of Shanghai Municipal Education Commission (no. 2019-01-07-00-02-E00066), National Science Foundation of China (grant numbers: 12202271, 52122110, 42076210), Program for Intergovernmental International S\&T Cooperation Projects of Shanghai Municipality, China (grant no. 22160710200), and the Oceanic Interdisciplinary Program of Shanghai Jiao Tong University (grant no. SL2020PT201).
KZ is also grateful for the computing resources at Amarel cluster provided through the Office of Advanced Research Computing (OARC) at Rutgers University, on which some of the simulations were carried out.
OB is supported by the Department of Energy Advanced Research Projects Agency-Energy Program award DE-AR0001186.

\bibliographystyle{unsrtnat}
\bibliography{references}  






\end{document}